\documentclass[11pt]{article}
\pdfoutput=1
\usepackage{aas_macros,jcappub,natbib,float,caption,comment,bm}
\usepackage{graphicx,epsfig}
\usepackage[dvipsnames]{xcolor}
\usepackage[utf8]{inputenc}

\def\be{\begin{equation}}
\def\ee{\end{equation}}
\def\ba#1\ea{\begin{align}#1\end{align}}
\newcommand{\vs}{\nonumber\\}

\let\oldv\v

\renewcommand{\v}[1]{\mathbf{#1}}

\newcommand{\vr}{\v{r}}	
\newcommand{\vk}{\v{k}}

\newcommand{\refeq}[1]{eq.~(\ref{eq:#1})}
\newcommand{\refeqs}[2]{eqs.~(\ref{eq:#1})--(\ref{eq:#2})}

\newcommand{\reffig}[1]{figure~\ref{fig:#1}}

\newcommand{\refsec}[1]{Section~\ref{sec:#1}}

\newcommand{\reftab}[1]{table~\ref{tab:#1}}

\renewcommand{\[}{\left[}
\renewcommand{\]}{\right]}
\renewcommand{\(}{\left(}
\renewcommand{\)}{\right)}

\newcommand{\hMpc}{~h^{-1}~{\rm Mpc}}
\newcommand{\ihMpc}{~h~{\rm Mpc}^{-1}}
\newcommand{\lya}{\text{Lyman-}\alpha}
\renewcommand{\d}{\delta}
\newcommand{\para}{\parallel}
\newcommand{\fnl}{f_{\rm NL}}
\newcommand{\dc}{\delta_c}

\newcommand{\dnl}{{\mathcal D}_{\rm NL}}

\newcommand{\err}{{\rm err}}

\title{Response approach to the squeezed-limit bispectrum: application to the correlation of quasar and Lyman-$\alpha$ forest power spectrum}
\author[a]{Chi-Ting Chiang,}
\author[b]{Agnieszka M. Cieplak,}
\author[c]{Fabian Schmidt,}
\author[b]{and An\oldv{z}e Slosar}
\affiliation[a]{C.N. Yang Institute for Theoretical Physics, Stony Brook University, Stony Brook, NY 11794, U.S.A.}
\affiliation[b]{Brookhaven National Laboratory, Blgd 510, Upton, NY 11375, U.S.A.}
\affiliation[c]{Max-Planck-Institut f\"ur Astrophysik, Karl-Schwarzschild-Str. 1, 85741 Garching, Germany}
\emailAdd{chi-ting.chiang@stonybrook.edu,acieplak@bnl.gov,fabians@mpa-garching.mpg.de,anze@bnl.gov}
\abstract{The squeezed-limit bispectrum, which is generated by nonlinear gravitational
evolution as well as inflationary physics, measures the correlation of three wavenumbers,
in the configuration where one wavenumber is much smaller than the other two. Since
the squeezed-limit bispectrum encodes the impact of a large-scale fluctuation on the
small-scale power spectrum, it can be understood as how the small-scale power spectrum
``responds'' to the large-scale fluctuation. Viewed in this way, the squeezed-limit
bispectrum can be calculated using the response approach even in the cases which do
not submit to perturbative treatment. To illustrate this point, we apply this approach
to the cross-correlation between the large-scale quasar density field and small-scale
Lyman-$\alpha$ forest flux power spectrum. In particular, using separate universe
simulations which implement changes in the large-scale density, velocity gradient,
and primordial power spectrum amplitude, we measure how the Lyman-$\alpha$ forest
flux power spectrum responds to the local, long-wavelength quasar overdensity, and
equivalently their squeezed-limit bispectrum. We perform a Fisher forecast for the
ability of future experiments to constrain local non-Gaussianity using the bispectrum
of quasars and the Lyman-$\alpha$ forest. Combining with quasar and Lyman-$\alpha$
forest power spectra to constrain the biases, we find that for DESI the expected
$1-\sigma$ constraint is ${\rm err}[f_{\rm NL}]\sim60$. Ability for DESI to measure
$f_{\rm NL}$ through this channel is limited primarily by the aliasing and instrumental
noise of the Lyman-$\alpha$ forest flux power spectrum. The combination of response
approach and separate universe simulations provides a novel technique to explore the
constraints from the squeezed-limit bispectrum between different observables.}

\begin{document}

\subheader{\rm YITP-SB-17-01}

\maketitle
\flushbottom

\section{Introduction}
\label{sec:introduction}
The universe today is highly nonlinear, which renders making analytical or
semi-analytical predictions a very hard problem even for the
heavily simplified case of a universe made up of collisionless dark
matter interacting through gravitational force alone. Luckily, the
gravitational force is the only relevant long-range force in the universe and
the curvature fluctuations are the only field that matters at the very
large scales in the universe. This allows us to make surprisingly
strong statements about properties of certain limits of correlators.
The basic picture is that of a ``separate universe.''

The separate universe picture posits that the evolution of a sufficiently
large patch of the universe riding a large-scale overdensity is the
same as that of a typical patch in a slightly overdense universe.
In other words, we are making the approximation that the effect
of a mode with sufficiently small wavenumber $k$ must approach that of the $k=0$ mode.
On one hand, this approximation can be used in formal results
\cite{Creminelli:2013mca,Peloso:2013zw,Dai:2015jaa,Hu:2016ssz,Hu:2016wfa}.
On the other hand, there is ample evidence, from numerical experiments,
that the information flow in the universe is strictly from the large
scales to small scales, rather than the other way around
\cite{Little:1991py,Melott:1992vp,Shandarin:2009ue,Schneider:2011wf,Einasto:2011eu,Pontzen:2015eoh}.
This means that in the sufficiently nonlinear regime, the small
scales will contain remnant information of the large-scale phases,
but the opposite is not true. This means that the separate universe
picture works very well even in the limits where it is not expected to
be formally exactly correct, i.e. it becomes an extremely good ansatz
for the dominant dynamics of the system. This motivates entire trunks
of research in the field and forms the basis of BAO reconstruction
methods \cite{Eisenstein:2006nk,Padmanabhan:2012hf} and the super-sample
variance effects \cite{Takada:2013bfn,Li:2014jra,Li:2014sga,Akitsu:2016leq}.

This separate universe picture can also be straightforwardly implemented
in cosmological $N$-body simulations, which is known as the separate universe
simulations \cite{Sirko:2005uz,Gnedin:2011kj,Li:2014sga,Wagner:2014aka}, to
study how the small-scale structure formation is affected by the large-scale
density environment. Separate universe simulations have allowed for detailed
studies on squeezed-limit $n$-point correlation function \cite{Wagner:2015gva},
the halo bias \cite{Lazeyras:2015lgp,Li:2015jsz,Baldauf:2015vio}
(see Ref.~\cite{Desjacques:2016} for a review),
the $\lya$ forest \cite{McDonald:2001fe,Cieplak:2015kra},
and of the effect of other sectors that possess non-gravitational forces such as
quintessence dark energy \cite{Chiang:2016vxa}.

In this paper, we add to the canon by presenting explicit recipes for making
predictions of the squeezed-limit bispectrum which couple one large-scale mode
from one field and two small-scale modes from another (which can, but is not
required to be the same as the first one). We consider the case most relevant
for actual surveys: we take into account not only the possible large-scale
density field, but also the redshift-space distortions and primordial non-Gaussianity
of the local-type, which we hope to measure with exactly this kind of measurement.
As a concrete example, we apply this method to the cross-correlation between
the large-scale quasar field and small-scale $\lya$ forest field and perform
a Fisher forecast for the ability of future experiments to constrain local
non-Gaussianity using this bispectrum.
Ref.~\cite{Hazra:2012qz} considered the
constraint on local primordial non-Gaussianity using the $\lya$ forest flux
bispectrum alone. However, as we shall argue in \refsec{conclusion}, the
cross-correlation is generally more robust regarding systematics than the auto-correlation, and so may be more appealing.

The rest of the paper is organized as follows.
In \refsec{theory}, we derive the relation between the power spectrum response
and the squeezed-limit bispectrum, and then use the small-scale $\lya$ forest
flux power spectrum and the large-scale quasar overdensity as an example.
In \refsec{sim}, we introduce the separate universe simulations that are used
to measure how the $\lya$ forest flux power spectrum responds to gravitational
evolution and primordial non-Gaussianity.  
We then use the measured responses as well as the Fisher matrix to explore
the constraining power on $\fnl$ from the quasar--$\lya$ forest squeezed-limit
bispectrum in \refsec{fnl}.
We conclude in \refsec{conclusion}.

\section{Theory}
\label{sec:theory}

\subsection{Warming up: large-scale biasing to first order}
\label{sec:warmup}
We start by re-deriving some standard results as a warm-up. Consider a
field $X$ (e.g. quasar overdensity field) with fluctuations on all
scales. Its overdensity $\d_X$ is defined as
\be
  X(\vr) = \bar{X}\[1+\d_X(\vr)\] \,.
\ee

Now, let us consider large patches of the universe and smooth $X$ on
such patches. In Fourier space, this procedure will suppress small-scale
fluctuations in $X$, but leave large-scale ones unaffected. The values
of large-scale fluctuations can now be Taylor expanded in the three
fields that are relevant at large scales. For small $k$, we have
\be
  \d_X(\vk) = b_\d^X \d(\vk) + b_\eta^X \eta(\vk) + b_\phi^X \phi(\vk) \,.
\label{eq:xexp}
\ee

The three relevant fields are 
\begin{itemize}
\item Matter density $\d(\vr) = \rho(\vr)/\bar{\rho}-1$.
\item Dimensionless gradient of the peculiar velocity along the line-of-sight
  $\eta=-H^{-1} \frac{\partial v_\para}{\partial r_\para}$, where $\v{v}$ and
  $\vr$ are both in comoving coordinate, and $H$ is the Hubble rate. This is
  the only scalar quantity that can affect the redshift-space distortions at
  linear order. Moreover, in linear theory it is given by $\eta(\vk) = f \mu^2 \d(\vk)$
  \cite{Kaiser:1987qv}, where $f$ is the growth rate and $\mu$ is the cosine
  of $\vk$ along the line-of-sight.
\item Primordial potential $\phi$. This term is present in cosmologies with
  local primordial non-Gaussianity \cite{Komatsu:2001rj}.  Specifically,
  $b^X_\phi\propto \fnl$, while $\phi(\vk)=M^{-1}(k)\d(\vk)$,
  where $M(k)=\frac{2}{3}\frac{D(a)}{H_0^2\Omega_m}k^2T(k)$ is the operator of
  Poisson equation with $D(a)$ and $T(k)$ being the linear growth and transfer
  function, respectively. The linear growth is normalized to the scale factor
  in the matter-dominated epoch.
\end{itemize}

While these three fields are not independent degrees of freedom, they have
different dependencies on $\vk$ on all scales. Therefore, averaged over a
given finite patch, their values are not related and we need to treat them
as three different fields.

The separate universe picture tells us that the biases are given by responses
of the mean field with respect to the change in a given large scale field:
\be
  b^X_\d = \frac{\partial \ln \bar{X}}{\partial \d} \,, \quad
  b^X_\eta = \frac{\partial \ln \bar{X}}{\partial \eta} \,, \quad
  b^X_\phi = \frac{\partial \ln \bar{X}}{\partial \phi} \,.
\ee
These derivatives, or biases, can be evaluated with analytical approximation
or numerical simulations. In particular, $b^X_\d$ is the standard linear bias
often denoted as $b_1$. For objects such as galaxies, we have a fairly good
analytical understanding of the bias parameters. For example, if the field is
conserved in response to stretching of the coordinates in the radial direction,
$b^X_\eta=1$, leading to the standard Kaiser result for the power spectrum.  
Similarly, using the peak-background split argument, one finds
$b^X_\phi = 2 \fnl \dc (b_\d - 1)$ \cite{Slosar:2008hx}, where $\fnl$ quantifies
the strength of the primordial non-Gaussianity and $\dc\approx1.686$ is the
density threshold. Note, however, that the analytical approximation fails for
other fields, such as the $\lya$ forest flux, whose evolution is highly nonlinear
and which is strongly affected by radiative transfer effects, and so numerical
simulations are necessary to evaluate their bias parameters.

In this paper we expand this formalism to the squeezed bispectrum and use it
to make predictions for cross-correlation between quasar field and the $\lya$
forest.

\subsection{Signal of the squeezed-limit bispectrum}
\label{sec:bi_sig}
Consider short-wavelength fluctuations $\d_Y$ in the presence of a long-wavelength
fluctuation $\d_X(\vk_3)$. The bispectrum is defined as
\be
 \langle\d_Y(\vk_1)\d_Y(\vk_2)\d_X(\vk_3)\rangle=(2\pi)^3\d_D(\vk_1+\vk_2+\vk_3)B_{YYX}(\vk_1,\vk_2,\vk_3) \,,
\label{eq:BYYX}
\ee
where $\delta_D$ is the Dirac delta function.  
Here we assume that the wavelength of $\d_X$ is much larger than that of
$\d_Y$, hence this bispectrum is in the so-called ``squeezed limit'', in
which $k_3\ll k_1\approx k_2$.

This squeezed-limit bispectrum can be regarded as the ``response'' of the
small-scale power spectrum $P_{YY}(k)$ to $\d_X$ \cite{Chiang:2014oga,Wagner:2015gva}.
Specifically, in the limit that $\d_X$ has infinitely long wavelength, the
power spectrum formed by the two small-scale modes is modulated by the
long-wavelength perturbation as
\be
 P_{YY}(\vk_S|\d_X)=\bar{P}_{YY}(\vk_S)\[1+\delta_{YY}(\vk_S|\d_X)\] \,,
\label{eq:PYY}
\ee
where $\delta_{YY}(\vk_S|\d_X)$ is the perturbation of the mean power spectrum $\bar{P}_{YY}$ due to $\d_X(\vk_L)$,
and we use the notation that $\vk_S \equiv \vk_1-\vk_3/2 \approx \vk_1$
for the small-scale mode and $\vk_L\equiv\vk_3$ for the long-wavelength
mode. In analogy with \refeq{xexp},
we expand $\d_{YY}$ including the large-scale tidal field as
\cite{Dai:2015jaa,Ip:2016jji,Akitsu:2016leq}
\be
 \d_{YY}(\vk_S|\d_X) = b^{YY}_\d(\vk_S)\d(\vk_L) + b^{YY}_\eta(\vk_S)\eta(\vk_L)
 + b^{YY}_\phi(\vk_S)\phi(\vk_L) + b^{YY}_K(\vk_S) \hat k_S^i \hat k_S^j K_{ij}(\vk_L) \,,
\label{eq:dYY_w_tidal}
\ee
where $K_{ij} \equiv (\partial_i\partial_j/\nabla^2 - \d^K_{ij}) \d$ is the
dimensionless scaled tidal field and $\d^K_{ij}$ is the Kronecker delta. Note
that $\eta = f \hat n^i\hat n^j K_{ij}$, where $\hat n$ is the line-of-sight
vector. The last term on the right-hand side of \refeq{dYY_w_tidal} corresponds
to the coupling of the large-scale tidal field to the small-scale power spectrum;
in fact, $b^{YY}_K$ is precisely analogous to $R_K$ defined in \cite{barreira/schmidt}. 
In principle the velocity gradient perpendicular to the line-of-sight and the
tidal terms beyond \refeq{dYY_w_tidal} would have dynamic impact on small scales,
but they only enter at the quadratic order in the long mode, and so are not relevant
at the order we are working in.

In this paper, we consider the response to $\eta$ as a projection effect,
i.e. stretching of the coordinate in the radial direction, whereas the response
to $K_{ij}$ is a dynamical effect, i.e. changing the evolution of the small-scale
power spectrum locally. Thus to compute $\d_{YY}$, we need to know the bias
parameters $b^{YY}_\d$, $b^{YY}_\eta$, $b^{YY}_\phi$, and $b^{YY}_K$ (which
are different from the large-scale biases $b^X_\d$, $b^X_\eta$, and $b^X_\phi$),
or equivalently how $\bar{P}_{YY}$ responds to the large-scale fluctuations.

To evaluate the bias parameters, numerical simulations are required
due to the nonlinear nature of the $\lya$ forest. We shall discuss this in more detail in \refsec{sim}. 
However, there are currently no results on simulations that include
a large-scale tidal field. Lacking such simulations, we will neglect this contribution
here. Since our main goal is to study the constraining power on the primordial
non-Gaussianity from the response of small-scale power spectrum to the large-scale
fluctuations, and we do not expect a significant degeneracy between the terms
$b^{YY}_\phi(\vk_S)\phi(\vk_L)$ and $b^{YY}_K(\vk_S) \hat k_S^i \hat k_S^j K_{ij}(\vk_L)$,
this approximation should not have a large impact. To be more robust, we shall
further only consider the angle-averaged squeezed-limit bispectrum, hence the
response to the large-scale tidal field will be suppressed (although not perfectly,
since the angle-average is performed in redshift space, not real space).

Taking the above approximation, we can rewrite $\d_{YY}$ as
\be
 \delta_{YY}(\vk_S|\d_X)=b_\d^{YY}(\vk_S)\d(\vk_L)+b_\eta^{YY}(\vk_S)\eta(\vk_L)+b_\phi^{YY}(\vk_S)\phi(\vk_L) \,,
\label{eq:dYY}
\ee
where the responses, or equivalently the biases of the $\lya$ flux power spectrum,
are given by
\be
 b_\d^{YY}(\vk_S)=\frac{\partial\ln\bar{P}_{YY}(\vk_S)}{\partial\d} \,, \quad
 b_\eta^{YY}(\vk_S)=\frac{\partial\ln\bar{P}_{YY}(\vk_S)}{\partial\eta} \,, \quad
 b_\phi^{YY}(\vk_S)=\frac{\partial\ln\bar{P}_{YY}(\vk_S)}{\partial\phi} \,.
\label{eq:bYY}
\ee
Combining \refeqs{PYY}{bYY}, we can write the squeezed-limit bispectrum formed
by $\d_Y-\d_Y-\d_X$ as
\be
 B^{\rm sq}_{YYX}(\vk_S,\vk_L)=\lim_{k_3\to0}B_{YYX}(\vk_1,\vk_2,\vk_3)
 =\bar P_{YY}(\vk_S)\sum_{A,B=\d,\eta,\phi}b^X_{A}b^{YY}_B(\vk_S) P_{AB}(\vk_L) \,.
\ee
In the following, we will omit the bar in $P_{YY}$ and the $\vk_S$ dependence
in the responses ($b^{YY}_B$) when no confusion can arise.
Note that there is no bar in $P_{AB}$ since it is the power spectrum
of the large-scale perturbations, which is approximately constant to the
local observer.

To leading order, the responses are independent of the wavenumber of $\d_X$,
as it is a uniform change in the local patch. We can use this result either
to predict or to measure the squeezed-limit bispectrum by studying how $P_{YY}$
responds to $\d_X$, or equivalently the correlation between $P_{YY}(\vk_S|\d_X)$
and $\d_X$. This correlation is similar to the position-dependent power spectrum
proposed in Ref.~\cite{Chiang:2014oga}, and then applied to the measurements of
the squeezed-limit three-point functions of BOSS DR10 CMASS sample \cite{Chiang:2015eza}
as well as the CMB lensing cross-correlating the $\lya$ forest flux power
spectrum \cite{Doux:2016xhg}.

Since we shall perform the Fisher analysis of this correlation in \refsec{fnl},
we need the variance of this signal.
As the correlation between $P_{YY}(\vk_S|\d_X)$ and $\d_X$ is equivalent
to the squeezed-limit bispectrum, its variance can be computed using the
formalism of bispectrum \cite{Scoccimarro:1997st,Sefusatti:2007ih}.
Specifically, in the Gaussian limit the error of $B_{YYX}$ is given by
\be
 \[\Delta B_{YYX}(\vk_1,\vk_2,\vk_3)\]^2=\frac{(2\pi)^3}{k_F^3}
 \frac{P_{YY,t}(\vk_1)P_{YY,t}(\vk_2)P_{XX,t}(\vk_3)}{N_T(\vk_1,\vk_2,\vk_3)} ~,
\label{eq:BYYX_var}
\ee
where $k_F=2\pi/(V_{\rm survey})^{1/3}$ is the fundamental frequency with
$V_{\rm survey}$ being the survey volume, $N_T$ is the number of triangles
with $\vk_1+\vk_2+\vk_3=0$, and the power spectrum with the subscript $t$
contains the signal and noise. Note that in \refeq{BYYX_var} we assume that
$k_1\neq k_3$.

As we discussed earlier in this section, our main goal is to study the
constraining power on the primordial non-Gaussianity from the squeezed-limit
bispectrum using the separate universe approach, so we shall consider the
bispectrum in which the long mode $\d_X$ is angle averaged to suppress the
contribution from the response of the large-scale tidal field, i.e.
\be
 B^{\rm sq}_{YYX}(\vk_S,k_L)
 =P_{YY}(\vk_S)\sum_{A,B=\d,\eta,\phi}b^X_{A}b^{YY}_BP_{AB}(k_L) \,.
 \ee
Since we consider the bispectrum in the squeezed limit, the number of triangles
can be approximated by the product of the number of long and short modes:
\ba
 \lim_{k_3\to0}N_T(\vk_1,\vk_2,k_3)\:&=N(\vk_S)N(k_L)
 =\frac{2\pi k_S^2\Delta k\Delta\mu_S}{k_F^3}\frac{2\pi k_L^2\Delta k}{k_F^3} \,,
\ea
where $\Delta k$ is the binning of the wavenumber and $\Delta\mu_S$ is the
binning of the cosine of $\vk_S$ along the line-of-sight. Note that we consider
$0\le\mu_S\le1$ to avoid double counting the Fourier modes, and if we take
$\Delta\mu_S=1$ we recover the standard result for the number of modes with
wavenumber $k_S$ (see e.g. \cite{Jeong:2008rj}). The variance of the squeezed-limit
bispectrum is thus
\be
 \[\Delta B^{\rm sq}_{YYX}(\vk_S,k_L)\]^2=
 \frac{(2\pi)^3}{k_F^3}\frac{P_{YY,t}^2(\vk_S)P_{XX,t}(k_L)}{N(\vk_S)N(k_L)} \,.
\ee

\subsection{Application to the Lyman-$\alpha$ forest flux power spectrum and quasar overdensity}
\label{sec:application}
Let us now consider a concrete example, in which $\d_Y=\d_F$ is the small-scale
$\lya$ forest flux fluctuation and $\d_X=\d_q$ is the long-wavelength quasar
density fluctuation. As described in \refsec{bi_sig}, we can treat the squeezed-limit
bispectrum as the cross-power spectrum of two fields: the large-scale quasar
overdensity field and the changes in the small-scale $\lya$ forest flux power
spectrum due to the long-wavelength fluctuations. Following \refsec{warmup}
and \refsec{bi_sig}, both fields trace the dominant large-scale fields, which
then generate the bispectrum we are investigating. On large scales we have
\ba
 \d_q(k_L)\:&=b^q_\d \d(k_L)+b^q_\eta \eta(k_L)+b^q_\phi \phi(k_L) \,, \vs
 \d_{FF}(\vk_S,k_L)\:& = b^{FF}_\d \d(k_L) + b^{FF}_\eta \eta(k_L)
 + b^{FF}_\phi \phi(k_L) \,.
\ea
Note that the responses of the small-scale $\lya$ forest flux power spectrum
to the long-wavelength fluctuations depend on the small-scale wavevector $\vk_S$,
so there is an implicit dependence in the bias parameters of $\lya$ forest flux
power spectrum fluctuation on $\vk_S$.

The halo (quasar) density field has been well studied \cite{Baldauf:2010vn,Tellarini:2016sgp},
and it has just one free parameter $b^q_\d$, with $b^q_\eta=1$ and
$b^q_\phi=2\fnl\d_c(b^q_\d-1)$ \cite{Dalal:2007cu,Matarrese:2008nc}.
The numerical value of $b^q_\d$ depends on the population of quasars
and typical halo mass hosting these quasars. For the $\lya$ flux power
spectrum, since the responses are highly nonlinear, the bias parameters
must be derived from simulations, and we shall discuss each components
in detail in \refsec{sim}.

The $\lya$ forest-quasar squeezed-limit bispectrum, in which quasars
serve as the long-wavelength mode, can be approximated by the correlation
between $P_{FF}$ and $\d_q$ as
\be
 B^{\rm sq}_{FFq}(\vk_S,k_L)
 =P_{FF}(\vk_S) \sum_{A,B=\d,\eta,\phi} b^q_A b^{FF}_B P_{AB}(k_L) \,,
\label{eq:iB_FFq}
\ee
where, in linear theory, the large-scale power spectra are just the linear
power spectrum $P_l$ weighted by the appropriate $k$ and redshift-space
distortion factors:
\ba
  P_{\d\d}(k_L)\:&= P_l(k_L) \,, \\
  P_{\d\eta}(k_L)\:&= \frac{f}{3} P_l(k_L) \,, \\
  P_{\d\phi}(k_L)\:&= M^{-1}(k_L) P_l(k_L) \,, \\
  P_{\eta\eta}(k_L)\:&= \frac{f^2}{5} P_l(k_L) \,, \\
  P_{\eta\phi}(k_L)\:&= \frac{f}{3} M^{-1}(k_L) P_l(k_L) \,, \\
  P_{\phi\phi}(k_L)\:&= M^{-2}(k_L) P_l(k_L)  \,,
\ea
where $M(k)=\frac{2}{3}\frac{D(a)}{H_0^2\Omega_m}k^2T(k)$ is the Poisson
operator that relates potential and density perturbations.
This is accurate if $k_L$ is sufficiently small so that linear theory applies.
The factors of $1/3$ and $1/5$ appear after angle-averaging over $\mu_L$.

The fiducial $\lya$ forest flux power spectrum of the short mode $\vk_S$
is given by \cite{McDonald:2001fe,Arinyo-i-Prats:2015vqa}
\be
 P_{FF}(\vk_S)=P_{FF}(k_S,\mu_S)
 =(b^F_\d)^2(1+\beta_F\mu_S^2)^2P_l(k_S)\dnl(k_S,\mu_S) \,,
\label{eq:arinyo}
\ee
where $b^F_\d$ is the flux bias, $\beta_F=f b^F_\eta/b^F_\d$ is the
redshift-space distortion parameter for the flux, and $\dnl(k_S,\mu_S)$
is the fitting formula for the small-scale nonlinearity. We reiterate
that $b^F_\d\neq b^{FF}_\d$: the former is the bias of the flux, given
by $\partial\ln\bar{F}/\partial \delta_L$, while the latter is the bias
of the flux power spectrum given by $\partial \ln P_{FF} / \partial \delta_L$.
In this paper, we shall adopt the fitting formula provided in Ref.~\cite{Arinyo-i-Prats:2015vqa}
as the fiducial model, which is described with more detail in \refsec{sim_s8}.

The remaining task is thus to compute how the $\lya$ forest power spectrum
responds to long-wavelength fluctuations, which will be discussed in \refsec{sim}.

\section{Simulations}
\label{sec:sim}

\subsection{Lyman-$\alpha$ forest flux power spectrum response to gravitational evolution}
\label{sec:sim_gv}
Since the dynamics governing the small-scale $\lya$ forest is highly nonlinear,
we have to rely on $N$-body simulations to study its responses with respect to
long-wavelength modes. A similar effect has been studied for the three-point
function of quasar lensing cross-correlating the $\lya$ flux power spectrum in
real space \cite{LoVerde:2010hj}, but we shall extend it to redshift space.

Ref.~\cite{Cieplak:2015kra} presented simulation measurements of the biases of
the $\lya$ forest flux fluctuations, which we will build on. In short, the fiducial
cosmology is flat $\Lambda$CDM with $\Omega_m=0.275$, $\Omega_\Lambda=0.725$,
$h=0.702$, and $\sigma_8=0.816$. The box size is $40\hMpc$ with $2\times1024^3$
particles for dark matter and gas. The hydrodynamics are carried out by GADGET-3
\cite{Springel:2005mi}, with Haardt and Madau UV background \cite{Haardt:1995bw}
and the simple QUICKLYA option for star formation without feedback.

The evolved small-scale density field $\d_S$  and the optical depth
field $\tau$ in real space are related by the 
Fluctuating Gunn Peterson Approximation (FGPA) given by
\be
\tau=-A(1+\d_S)^\alpha \,,
\label{eq:flux}
\ee
where $A$ is a constant (depending on the photoionization
rate, the gas temperature, and redshift) and $\alpha=2-0.7(\gamma-1)$ with $\gamma-1=d\ln\rho/d\ln T$
describing the temperature-density relation.
Since we shall use the parameters in
Ref.~\cite{Arinyo-i-Prats:2015vqa} as our fiducial $\lya$ forest flux power spectrum,
we choose $\alpha=1.58$ and $A=0.4$ to match the normalization of their $\lya$ forest
power spectrum at $z=2.6$. The redshift-space distortion is added to the optical
depth by
\be
 \tau_s(s_\para)=\int dr_\para~ \tau(r_\para)\delta_D\(s_\para-r_\para-H^{-1}v_\para\) \,,
\ee
where $s$ denotes the redshift-space coordinate and $v$ is the velocity field. Finally,
the thermal motion of neutral hydrogen gas would broaden the $\lya$ absorption,
and we introduce a Gaussian line broadening profile to the cross section as
\be
 \sigma_\alpha=\sigma_0\frac{c}{b\sqrt{\pi}}e^{-(\Delta v)^2}/b^2 \,,
\ee
where $\sigma_0$ is the cross section at rest, $c$ is the speed of light,
$b^2=(12.8~{\rm km~s}^{-1})^2\(\frac{T_0}{10^4~{\rm K}}\)(1+\d_S)^{\gamma-1}$
is the Doppler parameter, and $\Delta v$ is the velocity difference with respect
to the center of the absorption line. Finally, the redshift-space
optical depth can be used to infer the redshift-space flux:
\be
F(s_\para) = e^{-\tau_s(s_\para)} \,.
\ee

\begin{figure}[t]
\centering
\includegraphics[width=0.5\textwidth]{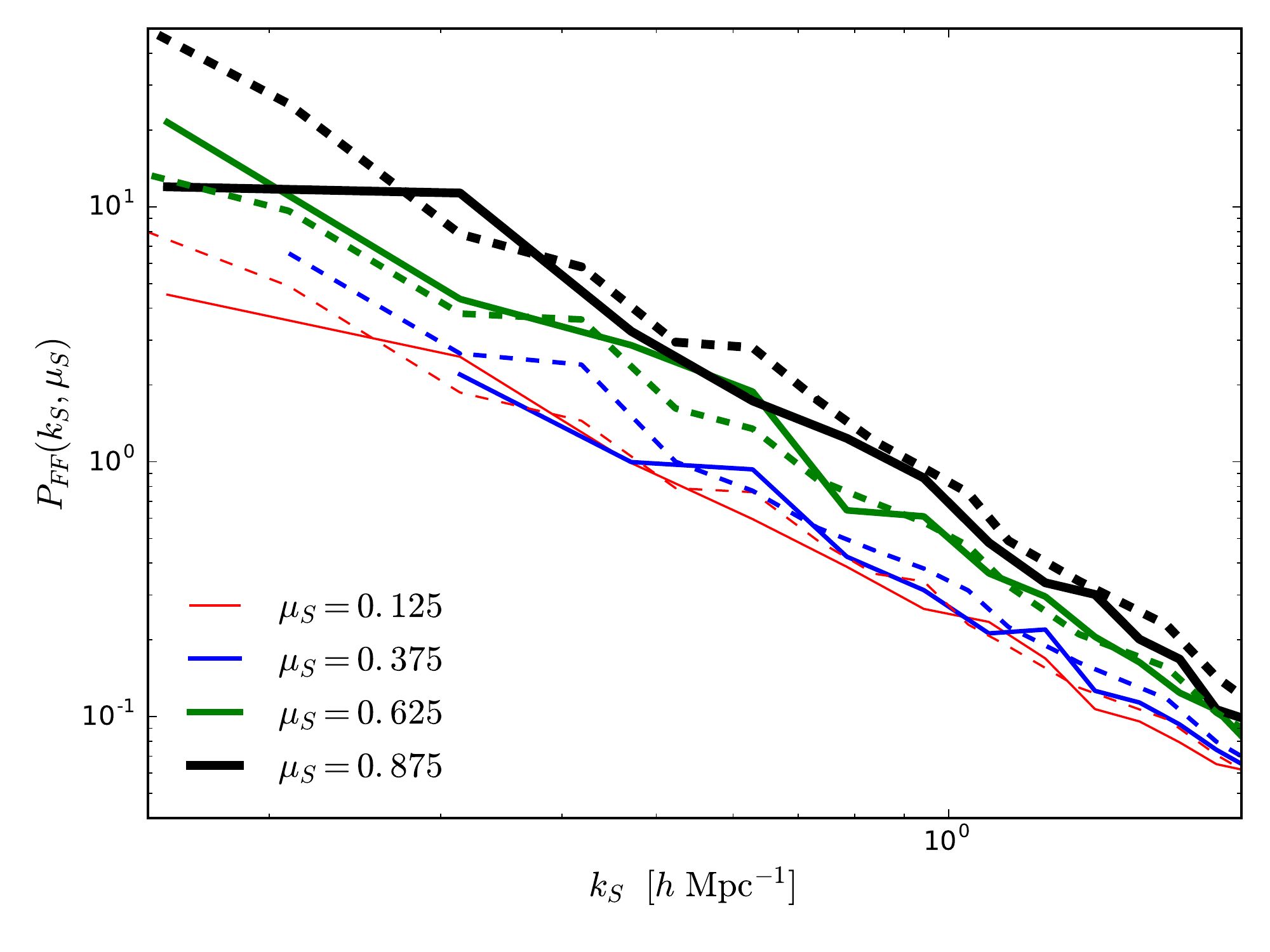}
\caption{The comparison of the mean $\lya$ forest flux power spectrum between
our prescription (solid) and the simulations in Ref.~\cite{Arinyo-i-Prats:2015vqa}
(dashed) for four different $\mu_S$.}
\label{fig:pffbar}
\end{figure}

In \reffig{pffbar} we compare the power spectrum of this prescription
to the simulations in Ref.~\cite{Arinyo-i-Prats:2015vqa}. We find that on
the scales of interest ($0.1\le k_S\le1\ihMpc$) the two power spectra are
in reasonable agreement. Since we are taking logarithmic derivatives,
these differences are unlikely to be very important.

In the following, we shall present the responses of the $\lya$ flux power spectrum to
large-scale $\d$ and $\eta$ in \refsec{sim_delta} and \refsec{sim_eta}, respectively.

\subsubsection{$\d$ response}
\label{sec:sim_delta}
To simulate the $\lya$ forest power spectrum in overdense and underdense regions,
or in the separate universe (SU), we set $\d(z=2.5)=\pm0.015$.
To the leading order the response is independent of the choice of $\d$,
as long as $\d\ll1$. To avoid numerical errors, however, we set $\d(z=2.5)$
to be large enough so that it is easy to extract the signal.
Due to the presence
of $\d$, the cosmology in SUs is affected, and the new parameters are listed in
table~1 of Ref.~\cite{Cieplak:2015kra}. Since the SU simulations are meant to
represent the overdense and underdense regions of the fiducial cosmology, there
are a few rescalings we need to perform on the density field as we apply the FGPA
prescription above, as well as on the $\lya$ forest flux power spectrum as we
compute the response.

The first rescaling is the density fluctuation. Since the mean densities in SUs
and the global universe are related by $\bar{\rho}_{\rm SU}=\bar\rho_{\rm G}\(1+\d\)$,
the mapping of the locally measured small-scale density fluctuation $\d_S$
to the global universe at leading order is
\be
 \d_S \rightarrow \d_S(1+\d)+\d \,.
\ee
We then use the remapped $\d_S$ for the FGPA prescription to compute the $\lya$
forest flux in SUs.

The second rescaling is the volume. To leading order, the scale factors in SUs
and the global universe are related by $a_{\rm SU}=a_{\rm G}(1-\d/3)$, with $\d$
now normalized for the appropriate redshift matching in our simulations as described
in the  Appendix of Ref.~\cite{Cieplak:2015kra}.
We would like to compare the observables at the same physical time and coordinate.
As the simulation box has the
same {\it comoving} volume, the {\it physical} volume would be changed by a factor
of $(1-\d)$. The power spectrum is normalized by the volume, hence to compare the
power spectrum in SUs with the one in the global universe we have to rescale it as
\be
 P_{FF}(k_S,\mu_S) \rightarrow (1+\d) P_{FF}(k_S,\mu_S) \,.
\ee

We finally rescale the wavenumber of the power spectrum measured in SUs.
The reason is identical to the volume rescaling: the power spectra in separate
universes are quantified in their own comoving coordinate with different scale factors,
so we need to rescale the wavenumbers to the comoving coordinate of the fiducial
universe for a fair comparison.
Since the wavenumber is proportional to the inverse of length, this results in a shift
in the power spectrum as
\be
 P_{FF}(k_S,\mu_S) \rightarrow P_{FF}(k_S,\mu_S) \[1+\frac{\d}{3}\frac{d\ln P_{FF}(k_S,\mu_S)}{d\ln k_S}\] \,.
\ee

\begin{figure}[t]
\makebox[\textwidth][c]{
 \includegraphics[width=1.05\textwidth]{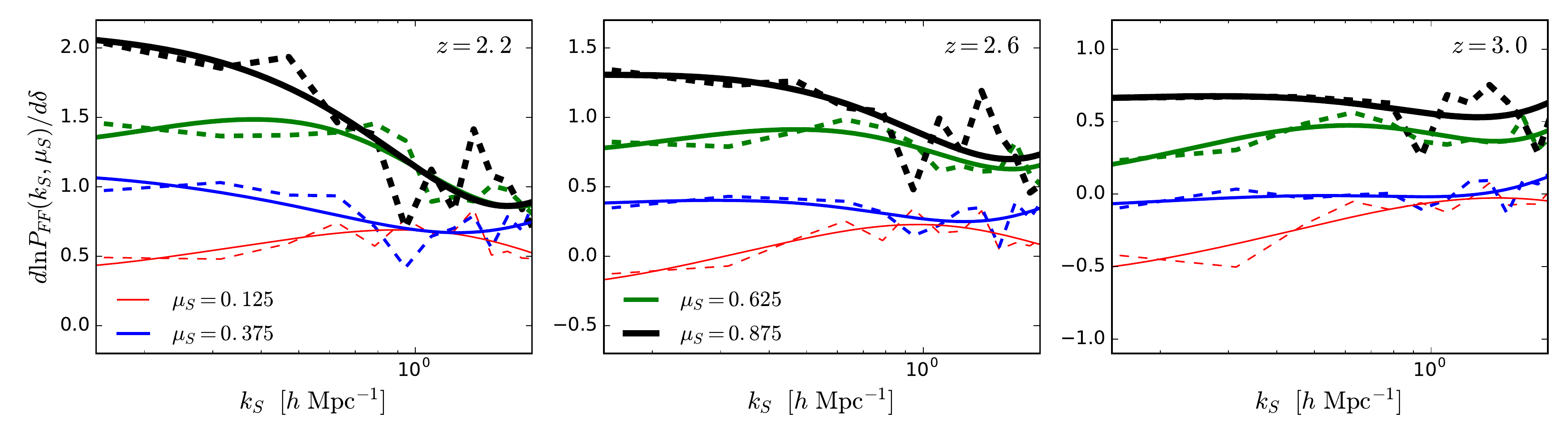}
}
\caption{The responses of the $\lya$ forest flux power spectrum to the
large-scale overdensity $\d$. The left, middle and right panels show
the response, $\partial\ln P_{FF}/\partial\d$ at $z=2.2$, 2.6, and 3.0,
respectively. The dashed lines are measurements of our SU simulations
with $\d(z=2.5)=\pm0.015$, whereas the solid lines are smoothed by the
Savitzky-Golay filter and then interpolated using the cubic spline (details
are described in the main text). The thin to thick lines show the responses
of different lines-of-sight with $\mu_S=0.125$, 0.375, 0.625, and 0.875.}
\label{fig:dlnpffddelta}
\end{figure}

We estimate the response of the $\lya$ forest flux power spectrum to $\d$ as
\be
 \frac{\partial\ln P_{FF}(k_S,\mu_S)}{\partial\d} = b^{FF}_\d
 = \frac{P_{FF}(k_S,\mu_S|\d_+)-P_{FF}(k_S,\mu_S|\d_-)}{P_{FF}(k_S,\mu_S)}\frac{1}{\d_+-\d_-} \,.
\ee
The dashed lines in \reffig{dlnpffddelta} show the measured $\partial\ln P_{FF}(k_S,\mu_S)/\partial\d$
at $z=2.2$ (left), 2.6 (middle), and 3.0 (right) from our SU simulations.
The thin to thick lines show the responses of different lines-of-sight with
$\mu_S=0.125$, 0.375, 0.625, and 0.875. Since the measurements are noisy,
we smooth them using the Savitzky-Golay filter \cite{NR} of window length
53 and poly order 8 and then apply the cubic spline to interpolate in $\ln k$-space.
We visually inspect the smoothing, in particular at $0.1\le k_S\le 1\ihMpc$,
which is used for the Fisher matrix calculation in \refsec{fnl}. Note that as
the main purpose of this paper is to study the constraint on $\fnl$, as long
as the dependences of scale, angle, and redshift of the response to $\delta$
are different from that to $\phi$, the choice of smoothing should have relatively
small impact on our result.
We find a mild redshift evolution, with a slightly larger response at lower
redshifts. We also find that on large scales ($k_S\lesssim1\ihMpc$) the
response along the line-of-sight is larger than that in the transverse
direction, which is most likely due to redshift-space distortions amplifying
the effect. We shall use the solid lines (smoothed responses) for the Fisher
forecast in \refsec{fnl}.

\subsubsection{$\eta$ response}
\label{sec:sim_eta}
Due to the peculiar velocities, the redshift- and real-space coordinates
in the radial direction are related through $s=r+v_{\para}/(aH)$.
In the presence of the large-scale velocity gradient, the redshift-space coordinate
transforms further as $s\to s-r\eta$, which is effectively a stretching in the radial
direction \cite{Cieplak:2015kra}. In order to conserve the total number of hydrogen
atoms, the optical depth must therefore change by $(1-\eta)^{-1}$. In addition,
due to this effective radial stretching we must also divide the thermal broadening
parameter by $(1-\eta)$. We apply these changes in the FGPA prescription to compute
the $\lya$ forest flux power spectrum, with $\eta=\pm0.01$.
Note that, as for $\d$, the response is independent of the
choice of $\eta$ at leading order.

\begin{figure}[t]
\makebox[\textwidth][c]{
 \includegraphics[width=1.05\textwidth]{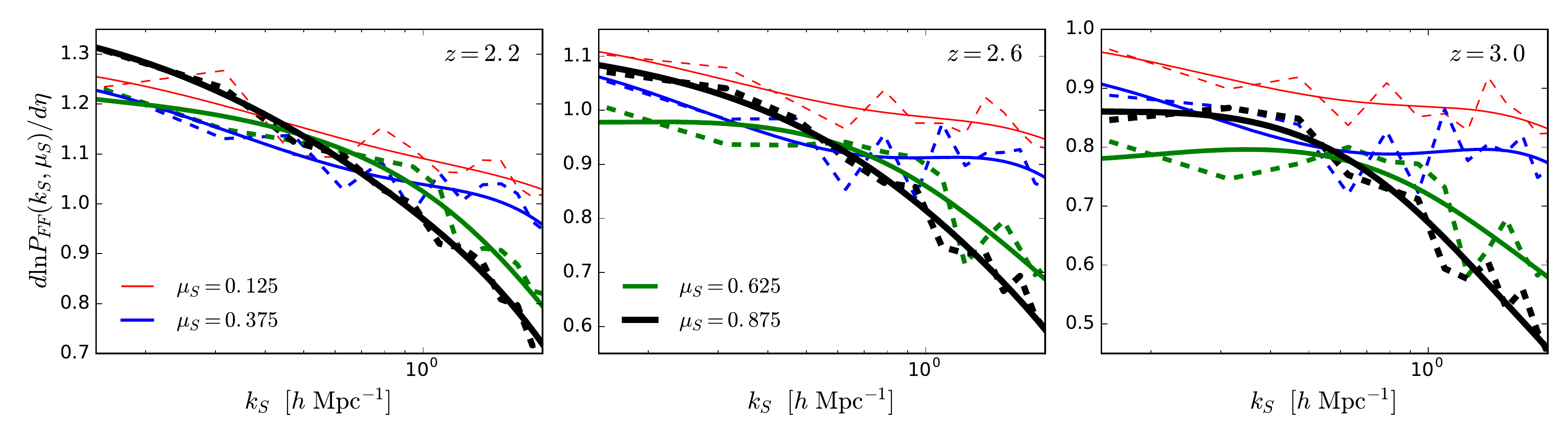}
}
\caption{Same as \reffig{dlnpffddelta}, but for the responses of the $\lya$
forest flux power spectrum to the large-scale velocity gradient $\eta$, i.e.
$\partial\ln P_{FF}/\partial\eta$.}
\label{fig:dlnpffdeta}
\end{figure}

We estimate the response of the $\lya$ forest flux power spectrum to $\eta$ as
\be
 \frac{\partial\ln P_{FF}(k_S,\mu_S)}{\partial\eta} = b^{FF}_\eta
 = \frac{P_{FF}(k_S,\mu_S|\eta_+)-P_{FF}(k_S,\mu_S|\eta_-)}{P_{FF}(k_S,\mu_S)}\frac{1}{\eta_+-\eta_-} \,.
\ee
The measurements of $d\ln P_{FF}(k_S,\mu_S)/d\eta$ from our SU simulations are
shown as the dashed lines in \reffig{dlnpffdeta}. As for the response to $\d$,
we smooth the raw measurements from simulations. However, as the scale-dependences
are different, for the response to $\eta$ we first smooth the measurements with
Savitzky-Golay filter of window length 255 and poly order 29, and then apply the
cubic spline to interpolate in $\ln k$-space.
We also visually inspect the smoothing at the scale of interest, i.e. $0.1\le k_S\le 1\ihMpc$.
The results are shown as the solid lines. We find that similar to $d\ln P_{FF}(k_S,\mu_S)/d\delta$,
$d\ln P_{FF}(k_S,\mu_S)/d\eta$ is larger at lower redshift. However, the response
to $\eta$ has stronger evolution in $\mu_S$. This is likely due to the higher
sensitivity of $P_{FF}(k_S,\mu_S)$ to $\eta$ than to $\d$. The solid lines
(smoothed responses) will also be used for the Fisher forecast in \refsec{fnl}.

\subsection{Lyman-$\alpha$ forest flux power spectrum response to primordial non-Gaussianity}
\label{sec:sim_s8}
While the $\lya$ forest flux power spectrum responds to large-scale $\d$ and
$\eta$ due to the nonlinear gravitational evolution, or equivalently non-zero
squeezed-limit bispectrum, in the presence of primordial non-Gaussianity it
will also be modulated by $\phi$. Specifically, the local-type primordial
non-Gaussianity changes $\sigma_8$ locally, and the $\lya$ forest flux power
spectrum is affected accordingly. Following the formalism in Ref.~\cite{Slosar:2008hx},
we can rewrite the derivative of $P_{FF}(k_S,\mu_S)$ with respect to $\phi$
into $\sigma_8$ as
\be
b^{FF}_\phi = \frac{\partial\ln P_{FF}(k_S,\mu_S)}{\partial\phi}
 =\frac{\partial\ln P_{FF}(k_S,\mu_S)}{\partial\sigma_8}\frac{\partial\sigma_8}{\partial\phi}
 =2\fnl\frac{\partial\ln P_{FF}(k_S,\mu_S)}{\partial\ln\sigma_8} \,.
\label{eq:phi2s8}
\ee
In Ref.~\cite{Arinyo-i-Prats:2015vqa}, a suite of simulations with different
$\sigma_8$'s has been performed, and so it is ideal to study how the $\lya$
forest flux power spectrum responds to the primordial non-Gaussianity.
Here we briefly describe the simulations and refer the readers to
Ref.~\cite{Arinyo-i-Prats:2015vqa} for more details.

The cosmological parameters for the simulations are $\Omega_m=0.3$,
$\Omega_\Lambda=0.7$, $\Omega_b=0.05$, $h=0.7$, and $n_s=1$. The three values for
$\sigma_8$ are 0.6396, 0.7581, and 0.8778. The box size is $60\hMpc$
with $2\times512^3$ particles (for dark matter and gas) and softening
length of $4~h^{-1}~{\rm kpc}$. The initial conditions are set up by
the Zel'dovich approximation at $z=49$, and the simulations are carried
out by GADGET-2 \cite{Springel:2005mi}. The $\lya$ forest flux power
spectrum is described by
\be
 P_{FF}(k_S,\mu_S)=(b^F_\d)^2(1+\beta_F\mu_S^2)^2P_l(k_S)\dnl(k_S,\mu_S) \,,
\ee
with the fitting formula accounting for the small-scale nonlinearity given by \cite{Arinyo-i-Prats:2015vqa}
\be
 \dnl(k,\mu)=\exp\left\lbrace q_1\Delta^2(k)\[1-\(\frac{k}{k_v}\)^{a_v}\mu^{b_v}\]
 -\(\frac{k}{k_p}\)^2\right\rbrace \,, \quad
 \Delta^2(k)=\frac{1}{2\pi^2}k^3P_l(k) \,,
\ee
where there are five fitting parameter $q_1$, $a_v$, $b_v$, $k_v$, and $k_p$.
The values of the bias and the fitting parameters for different values of
$\sigma_8$  can be found in table~8 and 9 of Ref.~\cite{Arinyo-i-Prats:2015vqa}.

We can estimate the response of the $\lya$ forest flux power spectrum
to $\sigma_8$ by
\be
 \frac{\partial\ln P_{FF}(k_S,\mu_S)}{\partial\ln\sigma_8}=\frac{\sigma_{8,0}}{P_{FF}(k_S,\mu_S|\sigma_{8,0})}
 \frac{P_{FF}(k_S,\mu_S|\sigma_{8,+})-P_{FF}(k_S,\mu_S|\sigma_{8,-})}{\sigma_{8,+}-\sigma_{8,-}} \,,
\label{eq:dlnpffslns8}
\ee
where $\sigma_{8,(-,0,+)}$ refers to 0.6396, 0.7581, and 0.8778 respectively.
Since Ref.~\cite{Arinyo-i-Prats:2015vqa} provides the fitting parameters as
well as the bias parameters of the $\lya$ forest flux power spectrum for
various redshifts and $\sigma_8$, we can thus numerically evaluate the
response with \refeq{dlnpffslns8}.
Note, however, that in Ref.~\cite{Arinyo-i-Prats:2015vqa} the same mean
flux is assumed for simulations with different $\sigma_8$'s. In reality,
different $\sigma_8$ would result in different long-wavelength density
perturbation. As the density fluctuation is nonlinearly related to the
flux, i.e. through \refeq{flux}, the large-scale mean flux would also
be different for simulations with different $\sigma_8$. Lacking enough
information to recover this effect, we ignore it in this paper but point
out that taking this effect into account will boost up the signal of $\fnl$.
As a result, our estimated constraint on $\fnl$ using the Fisher matrix
is likely to be conservative.

\begin{figure}[t]
\makebox[\textwidth][c]{
 \includegraphics[width=1.05\textwidth]{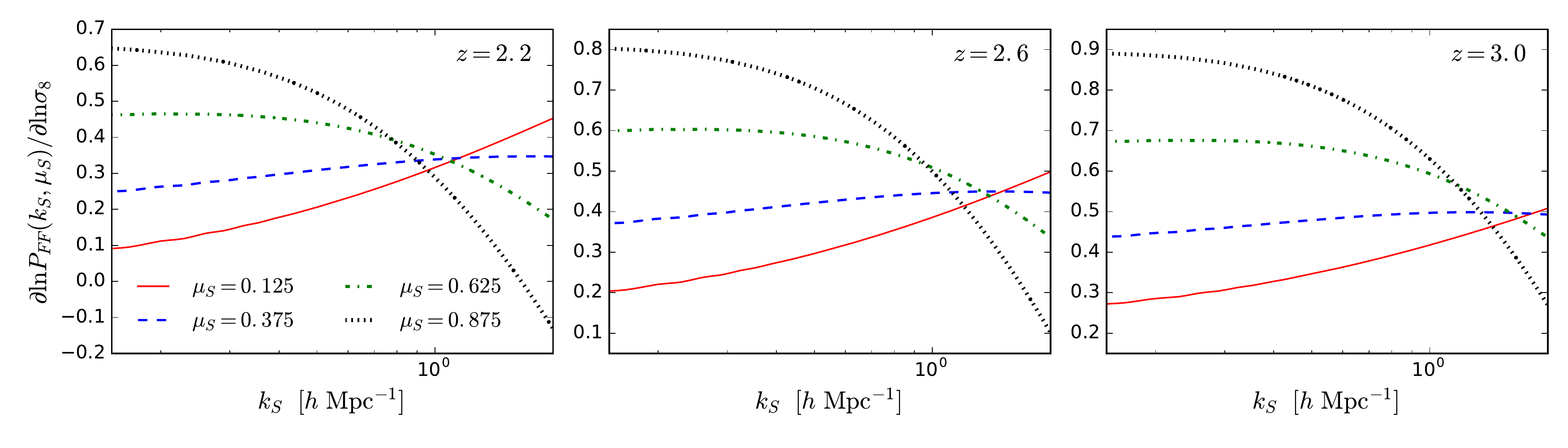}
}
\caption{The responses of the $\lya$ forest flux power spectrum to $\sigma_8$
(or equivalently the local-type primordial non-Gaussianity with \refeq{phi2s8})
at $z=2.2$ (left), 2.6 (middle), and 3.0 (right). The solid, dashed, dot-dashed,
and dotted lines show the response at $\mu_S=0.125$, 0.375, 0.625, and 0.875,
respectively.}
\label{fig:dlnpffdlns8}
\end{figure}

The left, middle, and right panels of \reffig{dlnpffdlns8} show the response
of the $\lya$ forest flux power spectrum to $\sigma_8$ at $z=2.2$, 2.6, and
3.0, respectively. Different line styles represent different lines-of-sight.
We find that opposite to the large-scale $\d$ and $\eta$, the $\lya$ forest
flux power spectrum responds stronger to the local-type primordial non-Gaussianity
at higher redshift. This trend is the same as the reduced matter squeezed-limit
bispectrum with local $\fnl$ (see e.g. figure~2.1 of Ref.~\cite{Chiang:2015pwa}).
Since the responses to gravitational evolution and primordial non-Gaussianity
have different redshift dependences, using the data from multiple redshifts
would help break the degeneracy between parameters and improve the constrains
on $\fnl$.

Another interesting but unfortunate fact is that while linear theory predicts
a scale-independent response $d\ln P_l/d\ln\sigma_8=2$, simulations yield
$d\ln P_{FF}(k_S,\mu_S)/d\ln\sigma_8<2$ on all scales and angles. This indicates
that the signal of the primordial non-Gaussianity from the response of the
$\lya$ forest flux power spectrum is smaller than that of the linear power
spectrum. We find that on large scales ($k\lesssim1\ihMpc$) the cancellation
is due to the Kaiser factor $(b^F_\d+b^F_\eta f\mu_S^2)^2$, and this cancellation
is weaker along the line-of-sight. The fitting formula $\dnl(k_S,\mu_S)$
damps the response along the line-of-sight on small scales ($k\gtrsim1\ihMpc$),
whereas it has negligible impact for the transverse direction on all scales.

A similar set of simulations has been performed in Ref.~\cite{McDonald:2001fe}
with different amplitudes of primordial spectrum $A_s$. Using their fitting
formula and parameters, we find a comparable result, i.e. the response is
larger for line-of-sight direction on large scale, while on small scales the
trend reverses, with the turning point at $\sim1\ihMpc$. The redshift dependence
of the response is also similar. Given the similarity between the two sets of
simulations, we use the more recent results for updated cosmological parameters
as well as better simulation resolution. As the response to $\d$ and $\eta$,
the results shown in \reffig{dlnpffdlns8} will be used for the Fisher forecast
in \refsec{fnl}.

\section{Expected constraint on primordial non-Gaussianity}
\label{sec:fnl}
One of the most exciting aspect of the large-scale structure is to constrain
the local primordial non-Gaussianity, from which we can learn whether inflation
is driven by one or multiple fields (see e.g. Ref.~\cite{Alvarez:2014vva}
for a review). Generally speaking, to acquire a competitive measurement of
the squeezed-limit bispectrum from large-scale structure, one needs both a
very large volume (to reach the squeezed limit), a sensitive probe (e.g. high
galaxy number density for galaxy bispectrum), and systematic robustness, since
the largest scales are typically the most contaminated in realistic large-scale
structure surveys. The cross-correlation between the large-scale quasar density
and small-scale $\lya$ forest flux power spectrum provides such an arena: (1)
the current and future quasar surveys have huge volume; (2) the cross-correlation
between these two tracers should be very clean. We will discuss this briefly
in \refsec{conclusion}.  

To explore the ability of constraining $\fnl$ by cross-correlating the
large-scale quasar density and small-scale $\lya$ forest flux power
spectrum, we use the Fisher matrix (see e.g. Ref.~\cite{Tegmark:1996bz}
for a review) to forecast the constraint. Specifically, the Fisher matrix
of $B_{FFq}^{\rm sq}(\vk_S,k_L)$ is given by
\be
 F_{\alpha\beta}=\sum_{k_S=k_{S,\rm min}}^{k_{S,\rm max}}\sum_{\mu_S=0}^1
 \sum_{k_L=k_{L,\rm min}}^{k_{L,\rm max}}
 \frac{1}{\[\Delta B_{FFq}^{\rm sq}(k_S,\mu_S,k_L)\]^2}
 \frac{\partial B^{\rm sq}_{FFq}(k_S,\mu_S,k_L)}{\partial p_\alpha}
 \frac{\partial B^{\rm sq}_{FFq}(k_S,\mu_S,k_L)}{\partial p_\beta} \,,
\label{eq:fisher}
\ee
where $p_\alpha\in(b_\d^q,b_\d^F,b_\eta^F,\fnl)$ are the parameters
of interest. The constraint on $p_\alpha$ as well as the correlation
between $p_\alpha$ and $p_\beta$ are then
\be
 \err\[p_\alpha\]=\sqrt{\(F^{-1}\)_{\alpha\alpha}} \,, \quad
 {\rm corr}\[p_\alpha,p_\beta\]=\frac{\(F^{-1}\)_{\alpha\beta}}{\err\[p_\alpha\]\err\[p_\beta\]} \,.
\label{eq:ifisher}
\ee

The signal of the squeezed-limit bispectrum is
\ba
 B^{\rm sq}_{FFq}(\vk_S,k_L)\:&=P_{FF}(\vk_S)P_l(k_L)
 \bigg\lbrace\[b^q_\d+2\fnl\dc\(b^q_\d-1\)M^{-1}(k_L)\]
 \[b^{FF}_\d+\frac{f}{3}b^{FF}_\eta+M^{-1}(k_L)b^{FF}_\phi\] \vs
 \:&\hspace{3.2cm}+\frac{f}{3}b^{FF}_\d+\frac{f^2}{5}b^{FF}_\eta
 +\frac{f}{3}M^{-1}(k_L)b^{FF}_\phi\bigg\rbrace \,,
\ea
and we take the partial derivatives of the signal with respect to the four
parameters with the equality that
\be
 \frac{\partial b^{FF}_\phi}{\partial\fnl}
 =\frac{\partial}{\partial\fnl}\[2\fnl\frac{\partial\ln P_{FF}(\vk_S)}{\partial\ln\sigma_8}\]
 =2\frac{\partial\ln P_{FF}(\vk_S)}{\partial\ln\sigma_8}=\frac{b^{FF}_\phi}{\fnl} \,.
\ee
For the error of the squeezed-limit bispectrum, we set $\Delta k=k_F$ and get
\be
 \[\Delta B_{FFq}^{\rm sq}(k_S,\mu_S,k_L)\]^2
 =\frac{(2\pi)^3}{k_F^3}P_{FF,t}^2(\vk_S)P_{qq,t}(k_L)\frac{k_F^4}{4\pi^2k_S^2k_L^2\Delta\mu_S} \,,
\ee
where the quasar power spectrum is
\be
 P_{qq}(k_L)=\[\(b^q_\d\)^2+\frac23 fb^q_\d+\frac15 f^2\]P_l(k_L) \,.
\ee
We follow Ref.~\cite{McDonald:2006qs} to compute the noise of the $\lya$ flux
power spectrum that includes both the aliasing noise and actual spectrograph
noise, whereas we assume that the noise of the quasar power spectrum is dominated
by the Poisson shot noise $P_{qq,\rm noise}=V_{\rm survey}/N_q$ with $N_q$ being
the number of quasars in the survey.

\begin{table}[t]
\centering
\begin{tabular}{ c c c c }
\hline \hline
 redshift & $V_{\rm survey}$ [$h^{-3}~{\rm Gpc}^3$] & $N_q/{\rm deg}^2/\Delta z$ & $N_q$ \\
\hline
 2.0-2.2 & 11.42 & 19 & 266,000 \\
 2.2-2.4 & 11.53 & 16 & 224,000 \\
 2.4-2.6 & 11.55 & 12 & 168,000 \\
 2.6-2.8 & 11.51 & 8  & 112,000 \\
 2.8-3.0 & 11.41 & 5  & 70,000 \\
\hline \hline
\end{tabular}
\caption{The survey parameters based on DESI for the Fisher forecast calculation.}
\label{tab:survey}
\end{table}

As a concrete example, we utilize the survey parameters of DESI \cite{Levi:2013gra,Aghamousa:2016zmz}
to numerically evaluate $F_{\alpha\beta}$. DESI will take spectra of quasars
at $2\le z\le3$ for $\lya$ forest absorption features across (at least)
14,000 square degrees, with the number density as a function of redshift given
in figure~3.17 of Ref.~\cite{Aghamousa:2016zmz}. In this paper we choose $\Delta z=0.2$,
and so there are five redshift bins. The key survey parameters are summarized
in \reftab{survey}, with the volume computed assuming the fiducial cosmology
in Ref.~\cite{Arinyo-i-Prats:2015vqa}, i.e. flat $\Lambda$CDM with $\Omega_m=0.3$
and $\Omega_b=0.05$. We assume the quasar bias to be $b_\d^q(z)=3.6 D(z=2.4)/D(z)$
\cite{Font-Ribera:2013fha}, where $D(z)$ is the linear growth. For the bias and
fitting parameters of the $\lya$ forest flux power spectrum, we adopt the values
in table~8 and 9 of Ref.~\cite{Arinyo-i-Prats:2015vqa}. We assume the fiducial
$\fnl$ to be zero, i.e. no primordial non-Gaussianity.

Since we are studying the constraining power for the squeezed-limit bispectrum
consisting of small-scale $\lya$ forest flux power spectrum and large-scale
quasar overdensity, we consider the range of scales to be $k_F\le k_L\le 0.05\ihMpc$
and $0.1\le k_S\le 1\ihMpc$, where the fundamental frequency $k_F=2\pi/V_{\rm survey}^{1/3}$
depends on the survey volume $V_{\rm survey}$ in various redshift bins. We consider
four lines-of-sight for the $\lya$ forest flux power spectrum, i.e. $\Delta\mu_S=0.25$
for $\mu_S=0.125$, 0.375, 0.625, and 0.875 as the results shown in \refsec{sim}.
The Fisher matrix is thus the summation of all scales, lines-of-sight, and
redshifts for \refeq{fisher}.

For one redshift bin, there are three bias parameters ($b_\d^q$, $b_\d^F$, and
$b_\eta^F$). Note that while we assume that the fiducial bias parameters at
different redshifts are related by the linear growth, in the Fisher matrix we
still conservatively treat them as independent parameters. In total there are
15 bias parameters, and so the Fisher matrix has the dimension of 16 including $\fnl$.

Using the inverse Fisher matrix, we compute the 1-$\sigma$ constraints on the
parameters as well as their correlation through \refeq{ifisher}. We find
$\err[\fnl]=77$ from using the squeezed-limit bispectrum of the small-scale
$\lya$ forest flux power spectrum and the large-scale quasar overdensity alone
for DESI survey parameters. To better understand what dominates the constraint
on $\fnl$, we artificially set the noise of quasar and $\lya$ forest flux power
spectrum to zero. In the absence of the shot noise of the quasar power spectrum,
we obtain $\err[\fnl]=46$, whereas in the absence of the noise of the $\lya$
forest flux power spectrum, we have $\err[\fnl]=1.2$. In the absence of both
quasar and $\lya$ forest flux noise, i.e. the sample variance limit, the constraint
becomes $\err[\fnl]=0.7$. It is thus clear that the noise of the $\lya$ forest
flux power spectrum dominates the constraint on $\fnl$. Specifically, we find
that at $z=2.2$ the signal-to-noise ratio of the $\lya$ forest flux power spectrum
per $k$ mode is of order $10^{-1}$ at $k_S=0.1\ihMpc$ and $10^{-3}$ at $k_S=1\ihMpc$.
The signal-to-noise ratio is even smaller at higher redshift.

On top of the constraint on $\fnl$, we also find that the constraints on the biases
are poor due to the high degeneracy: the typical absolute values of cross-correlation
coefficients between $b_\d^q$, $b_\d^F$, and $b_\eta^F$ in one redshift bin are all
greater than 0.99. This is because only the three-point function is used to constrain
the parameters. To break the degeneracy between biases, we add both the quasar and
$\lya$ forest flux power spectra into the Fisher calculation to constrain $b_\d^q$,
$b_\d^F$, and $b_\eta^F$ (but not $\fnl$). We use the same scales for the power
spectra as for the squeezed-limit bispectrum, and assume no covariance between the
three measurements. We find that including the power spectra largely reduces the
correlation between $b_\d^q(z)$ and $(b_\d^F,b_\eta^F)$; $b_\d^F$ and $b_\eta^F$ are
still highly anti-correlated due to the lack of lines-of-sight. Most interestingly,
breaking the degeneracy between biases also improves the constraint on $\fnl$ by
30\%, to $\err[\fnl]=56$.

In principle, there is also signal for $\fnl$ from the quasar scale-dependent bias,
i.e. in $b^q_\phi$. If this is included in the Fisher analysis, we obtain
$\err[\fnl]=4.6$, which indicates that the constraint on $\fnl$ is dominated by the
quasar scale-dependent bias. This is due to the low signal-to-noise ratio of the
quasar-$\lya$ forest squeezed-limit bispectrum, compared to the quasar scale-dependent
bias. Note that the advantage of the cross-correlation is that the signal suffers
less observational systematics than the auto-correlation, and so the result is
generally cleaner. Moreover, since different observables suffer different systematics,
it is useful to constrain $\fnl$ using multiple measurements to confirm the result.

We would like to caution the readers that some astrophysical effects
are neglected in our calculation. For example, due to the clustering
of galaxies and quasars, on scales larger than the mean-free-path of
the ionizing photons ($\sim350$ Mpc) the UV background is not uniform,
and so the ionization equilibrium in regions with greater separations
would depend on the density of the local regions
\cite{Pontzen:2014ena}. This introduces two new bias parameters
  $b_\Gamma^F$ and $b_\Gamma^{FF}$ which correspond to the response of the
  mean flux and power spectrum to fluctuations in the photoionization
  background. The actual calculation of the cross-correlation between
  the $\lya$ forest flux power spectrum and the large-scale UV
  fluctuation is beyond the scope of this paper.  However, we expect
that such an effect would have different scale-dependences compared to
the squeezed-limit bispectrum due to local primordial non-Gaussianity,
therefore if the effect is correctly modeled then the constraint on
$\fnl$ should not be biased or degrade, as for the constraint on dark
energy using BAO from $\lya$ forest flux power spectrum
\cite{Pontzen:2014ana}.

\section{Conclusions}
\label{sec:conclusion}
In this paper, we show that the squeezed-limit bispectrum $B_{YYX}^{\rm sq}$,
where $Y$ is the small-scale modes and $X$ the large-scale mode, can be computed
as the response of the small-scale power spectrum $P_{YY}$ to the large-scale
$\d_X$ fluctuation. This is similar to the position-dependent power spectrum
in which one measures the correlation between $P_{YY}$ and $\d_X$ \cite{Chiang:2014oga}.
Using the small-scale $\lya$ forest flux power spectrum and the large-scale
quasar overdensity as an example, we predict their squeezed-limit bispectrum
to be the responses to the large-scale density fluctuation $\d$, the large-scale
velocity gradient $\eta$, and the local-type primordial non-Gaussianity $\phi$.

Since the responses of the $\lya$ forest flux power spectrum are highly nonlinear,
we measure them from separate universe simulations. Specifically, a long-wavelength
$\d$ can be equivalently understood as modifying the local cosmology, and we can
directly simulate the structure formation in such an environment by adjusting the
local cosmological parameters. In the presence of $\eta$, the redshift-space coordinate
is stretched in the radial direction, and we thus rescale the small-scale $\lya$ forest
flux power spectrum in the line-of-sight direction. Finally, in the presence of local
primordial non-Gaussianity, the local $\sigma_8$ is modulated by the primordial potential
$\phi$, and we apply the fitting formula and parameters in Ref.~\cite{Arinyo-i-Prats:2015vqa}
to model the small-scale $\lya$ forest flux power spectrum with different $\sigma_8$.
The response of the small-scale $\lya$ forest flux power spectrum to the large-scale
fluctuations can thus be estimated by taking numerical derivatives with respect to
$\d$, $\eta$, and $\phi$. With the responses measured from the separate universe
simulations, we can thus numerically evaluate the squeezed-limit bispectrum of the
$\lya$ forest flux power spectrum and quasar overdensity (see Ref.~\cite{Chiang:2017jnm}
for the squeezed-limit bispectrum where the responses of the small-scale power spectrum
to the large-scale fluctuations can be computed perturbatively and analytically).

We then explore the constraining power of the local primordial non-Gaussianity
$\fnl$ with the squeezed-limit bispectrum of $\lya$ forest flux power spectrum
and quasar overdensity. We apply the response approach to predict the signal
of the bispectrum and the Fisher matrix to test the observability of $\fnl$.
In principle, this measurement should be systematically very clean: the selection
function, reddening, point spread function, and similar effects that plague galaxy
surveys at large scales and so contaminate quasar catalogs will not systematically
correlate with measurements of the small-scale $\lya$ forest flux power spectrum.
If fewer quasars are detected, this does not affect the measurements of the small-scale
power spectrum in the forest, but only its noise. Similarly, effects that are slowly
varying with observed frequency (reddening, loosing of photons due to point spread
function smearing, etc) are not affecting the size of small-scale fluctuations
relative to the mean. This makes this measurements considerably more appealing
than, for example, the quasar auto-bispectrum. On the other hand, there are astrophysical
systematics that could enter, via other, non-gravitational sources of large-scale
fluctuations in the Universe, such as those arising from photoionization rate
and temperature fluctuations \cite{Pontzen:2014ena}. But these must still respect
physical constraints which results in different scale dependencies (on sufficiently
large scales) that can be isolated and separated from primordial non-Gaussianity.

We find that for DESI the expected constraint is $\err[\fnl]\sim80$ using the
squeezed-limit bispectrum alone, and the result is dominated by the error of
the $\lya$ forest flux power spectrum. If we include both the quasar and $\lya$
forest flux power spectra to break the degeneracy between biases, the constraint
on $\fnl$ improves to $\err[\fnl]\sim60$. Note that while the effect of the
large-scale tidal field is neglected, we do not expect the constraint on $\fnl$
to worsen much if the tidal field is included, as it has different dependences
on scale and angle from the local primordial non-Gaussianity, and so the degeneracy
should be weak. Similar argument applies to other astrophysical effects that may
be ignored in our calculation. However, we note that our derivatives with 
respect to $\sigma_8$ variation are calculated at a fixed mean flux, since the
authors of Ref.~\cite{Arinyo-i-Prats:2015vqa} do not provide sufficient information
to correct for their renormalization of flux. While this underestimates the real
effect, we also know that the continuum fitting procedure typically results in
the power spectrum being measured around the local value of the mean flux. This
subtlety might affect our sensitivity estimate, but is unlikely to be dominant.

While we use the $\lya$ forest flux forest power spectrum and the quasar
overdensity as a concrete example to predict their squeezed-limit bispectrum
using the response approach, this approach can be applied to other observables
to forecast the signal of the squeezed-limit bispectrum. Especially, the
separate universe simulations are very powerful to estimate the responses
of the small-scale structure formation to the large-scale fluctuations,
the response approach is thus useful to compute the squeezed-limit bispectrum
of various observables for future observations.

Another application for our $\lya$ forest flux power spectrum response calculation
is to compare with the measurement of the cross-correlation between CMB lensing and
$\lya$ forest flux power spectrum recently presented in Ref.~\cite{Doux:2016xhg}.
In that work, a cross-correlation is done in real space, between the CMB lensing
convergence field $\kappa$ and the local measurements of the power spectrum. The
authors quantify their results in terms of the effective $b_2^{\rm eff}$ parameter,
which encodes the excess of the $\lya$ flux power spectrum response with respect to
that of the linear field. Since in our simulations, the flux power spectrum response
is always lower than that of the linear field, for all scales and lines-of-sight,
this would imply $b_2^{\rm eff}<0$, while Ref.~\cite{Doux:2016xhg} find $b_2^{\rm eff}=1.16\pm0.53$.
There are many reasons for this discrepancy that could include a surprisingly large
inaccuracies in our simulations or more likely the presence of unaccounted physical
effects, such as damped $\lya$ systems or temperature or photoionization rate fluctuations.
Since the presented measurement is a configuration-space measurement evaluated at
zero separation, it is also possible that our approximation of a squeezed limit
breaks down (i.e. there is significant contribution from other triangles).
We leave a detailed comparison with this measurement for future work.

\acknowledgments
We would like to thank Cyrille Doux and Emmanuel Schaan for explaining the
details of the measurement in Ref.~\cite{Doux:2016xhg}.
We would also like to thank Eiichiro Komatsu and Marilena LoVerde for helpful
discussions, as well as the referees for useful comments on the draft.
CC is supported by grant NSF PHY-1620628.  
FS acknowledges support from the Marie Curie Career Integration Grant (FP7-PEOPLE-2013-CIG) ``FundPhysicsAndLSS,''
and Starting Grant (ERC-2015-STG 678652) ``GrInflaGal'' from the European Research Council.


\bibliography{references}
\end{document}